\documentclass
[letter,twocolumn]{jpsj3} 
\usepackage{mathrsfs}

\title{Kondo Effect of a Vibrating Magnetic Impurity}
\author{Satoshi YASHIKI\thanks{E-mail address: syashiki@issp.u-tokyo.ac.jp}, 
        Shunsuke KIRINO and Kazuo UEDA}
\inst{Institute for Solid State Physics, University of Tokyo, Kashiwa 277-8581, Japan}
\abst{A generalized Anderson model for a magnetic ion in a harmonic potential is formulated. 
The model is investigated by the numerical renormalization group(NRG) method. 
In addition to the conventional $s$-wave screening, the model exhibits phonon assisted $p$-wave 
Kondo effect as well as Yu and Anderson type Kondo effect. It is shown that the $s$-wave Kondo and the
Yu-Anderson Kondo belong to the same fixed point. 
At the boundary between the $s$-wave and $p$-wave Kondo regions line of fixed points of 
the two channel Kondo effect is identified.}
\kword{Kondo effect, impurity Anderson model, electron-phonon coupling, 
       numerical renormalization group, two channel Kondo effect}
\begin{document}
\maketitle

Filled skutterudite compounds RT$_{4}$X$_{12}$ (R=
rare earth or alkaline earth element; 
T=Fe, Ru, Pt, or Os; X=P, As, Ge, or Sb) 
are characterized by their specific structures which 
involve a network of cages filled by guest ions. When 
the radius of the guest ion is smaller than the diameter of 
the cage, the guest ion vibrates with larger amplitude 
and smaller frequency than conventional localized modes.  
Such anharmonic local vibrations are 
referred to as rattling modes and may lead to novel 
phenomena of the coupled electron-phonon systems.  
Recent experiments on SmOs$_{4}$Sb$_{12}$ show a large electronic 
specific heat coefficient linear in $T$, which is robust 
against magnetic field\cite{mag_robust}.  One possible scenario of 
the unusual behavior is the effect of 
strong electron-phonon coupling.

When a magnetic ion couples with conduction electrons we expect 
Kondo effect due to the localized moment. On the other hand, it has been shown by 
Yu and Anderson(YA) that when conduction electrons couple strongly 
with ionic vibrations a different type of Kondo effect is 
expected\cite{Yu_Anderson}. In this Letter we will study 
the interplay between the conventional Kondo effect and 
the Yu-Anderson type one. Concerning the effects of coupling 
between a magnetic ion and ionic vibrations, Hotta has studied effects of 
anharmonicity in the Holstein-Anderson model\cite{Hotta}.
A lattice version of the Holstein-Anderson model is also studied\cite{mitsumoto_ono}.
In this Letter, we will 
discuss effects of transverse vibrations on the Kondo effect rather than the 
breathing type vibrations.

Suppose electron orbitals of a magnetic ion are described by the $\ell$-spherical-wave 
functions. Then the hybridization with the conduction electrons is given by overlap integrals 
between the $\ell$-wave localized orbitals and plane waves. To include the effect of 
vibrations the origin of the localized orbitals is shifted by $\mathbf{Q}$ which is the coordinate 
of the ion position. We can expand the overlap integral with respect to $\mathbf{Q}$. 
In the zeroth order the localized $\ell$-spherical-waves hybridize with the $\ell$-partial-waves 
of conduction electrons. In the first order of $\mathbf{Q}$, they hybridize 
with the $\ell+1$ and $\ell-1$ partial waves. In the present Letter we will concentrate 
on the simplest case of s-wave localized orbital. Then the total Hamiltonian 
up to the order of $\mathbf{Q}$ is given by the sum of
\begin{align}
    \mathit{H}_{\text{ion}}
=&  \hbar\omega\sum_{i}a^{\dagger}_{i}a_{i}\label{ion}
  + \varepsilon_{f}\sum_{\sigma}f^{\dagger}_{\sigma}f_{\sigma} + 
    Uf^{\dagger}_{\uparrow}f_{\uparrow}f^{\dagger}_{\downarrow}f_{\downarrow},\\
    \mathit{H}_{\text{c}}
=&  \sum_{k\sigma} \varepsilon(k)[c^{\dagger}_{0 \sigma}(k)c_{0 \sigma}(k)
   +\sum_{i}c^{\dagger}_{1i\sigma}(k)c_{1i\sigma}(k)],\\
    \mathit{H}_{\text{hyb}} 
=& \sum_{k\sigma}V_{0}\bigl[c^{\dagger}_{0\sigma}(k)f_{\sigma}
                                            +(h.c.)\bigr]\notag\\
   &+\sum_{k\sigma}\sum_{i}V_{1}
            \bigl[c^{\dagger}_{1i\sigma}(k)f_{\sigma}+(h.c.)\bigr](a_{i}+a^{\dagger}_{i}),
\end{align}
where $i$=$x, y, z$ represent three components of displacement vector as well as the three components 
of $\ell$=$1$ partial waves in real representation. In this paper phonons are assumed to be harmonic, 
and in Eq.(\ref{ion}) $a_{i}$ is an annihilation operator for the $i$-component phonon and $\omega$ 
is the frequency of the phonons.  Annihilation and creation operators for the localized electrons are 
described by $f_{\sigma}$ and $f^{\dagger}_{\sigma}$, $\epsilon_{f}$ is the energy of 
the localized orbital and $U$ the Coulomb interaction. $V_{0}$ is the matrix element 
of the hybridization of the $s$-wave channel and concerning the $p$-wave channel hybridization 
takes place assisted with phonons with the matrix element $V_{1}$.

\begin{figure}[!ht]
   \begin{center}
      \resizebox{60mm}{!}{\includegraphics{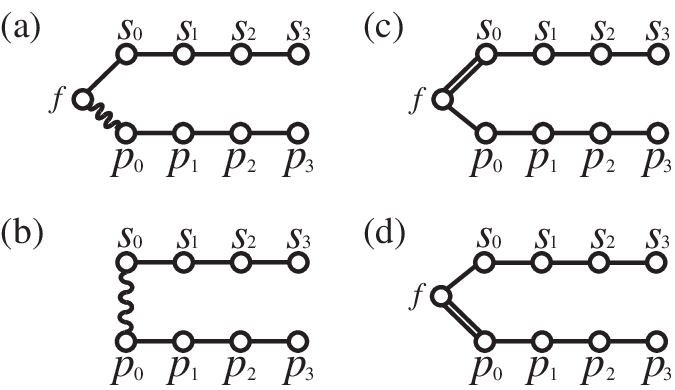}} 
   \end{center}
   \caption{(Color online)
(a) NRG scheme of the present model. 
The wavy line represents hopping processes accompanied with phonon creation or absorption. 
(b) The NRG scheme for the original Yu-Anderson model. 
(c) Low energy fixed point of the conventional $s$-chK.
(d) Low energy fixed point of the $p$-chK.}
   \label{site}
\end{figure}

To investigate properties of the model we first simplify the model to include vibrations of 
only one direction, say $i$=$x$.  We use the NRG method to analyze the model in detail\cite{Wilson,kri}. 
In the NRG algorithm we assume a constant density of states $\rho$=$1/2D$ with band width of $2D$ and 
discretize it in logarithmic energy scales controlled by $\Lambda$ for both the $s$-wave and $p$-wave channels.  
Then the present problem is mapped to the set of Hamiltonians where the localized orbital 
is coupled with two Wilson chains corresponding to the $s$-wave and $p$-wave channels
\begin{align}
 \mathit{H_{N}}
&=\varLambda^{\frac{N-1}{2}}\Biggl\{
 \frac{\widetilde{U}}{2}\biggl(\sum_{\sigma}f^{\dagger}_{\sigma}f_{\sigma}-1\biggr)^{2} 
  +\hbar\widetilde{\omega}a^{\dagger}a\notag\\
&+\sum^{N-1}_{n=0,\sigma}\varLambda^{-\frac{n}{2}}
\xi_{n}\bigl[s^{\dagger}_{n,\sigma}s_{n+1,\sigma}
  +p^{\dagger}_{n,\sigma}p_{n+1,\sigma}+(h.c.)\bigr]\notag\\
 &+\sum_{\sigma}\bigl[  \widetilde{V_{0}}s^{\dagger}_{0,\sigma}f_{\sigma}
                       +\widetilde{V_{1}}p^{\dagger}_{0,\sigma}f_{\sigma}(a+a^{\dagger})+(h.c.)\bigr]\Biggr\},
\end{align}
where the quantities with tilde are the parameters modified by the logarithmic discretization.
Figure.\ref{site}(a) shows the present system schematically.
We assume the electron-hole symmetry and thus the average number 
of electrons is kept to be one per site.

We first discuss several limiting cases. When the Coulomb interaction is strong 
electrons in the magnetic ion behave as a magnetic moment. When $V_{0}$ is stronger than $V_{1}$, 
the magnetic moment is quenched mainly through the coupling with the $s$-channel conduction electrons 
as temperature is lowered. This situation is a case of standard Kondo effect and will 
be referred to as $s$-channel Kondo ($s$-chK) regime in this Letter. On the other hand, when $V_{1}$ is dominant 
over $V_{0}$, the magnetic moment is screened mainly by the $p$-wave channel and this regime 
will be called $p$-channel Kondo ($p$-chK) scheme. 

When both the Coulomb interaction and the phonon assisted hybridization 
with the $p$-wave channel are not important the system may be treated as a renormalized Fermi(RF) chain. 
An interesting case in the weak Coulomb interaction regime is realized when the electron-phonon coupling 
$V_{1}$ is strong. It is instructive to consider the two-site problem which is 
described by the following Hamiltonian:
\begin{align}
 \mathit{H}_{\text{$2$-site}}
= \widetilde{V_{1}}
  \sum_{\sigma}\bigl[p^{\dagger}_{0\sigma}f_{\sigma}+(h.c.)\bigr](a+a^{\dagger})
 +\hbar\widetilde{\omega}a^{\dagger}a
\end{align}
This problem is solved easily by the canonical transformation\cite{Lang-Firsov}, $e^{iS}$ with
$S=i\lambda\sum_{\sigma}(-A^{\dagger}_{0\sigma}A_{0\sigma}
                         +B^{\dagger}_{0\sigma}B_{0\sigma})(a-a^{\dagger})$
where $A_{0\sigma}$ and $B_{0\sigma}$ are the annihilation operator for 
the antibonding and bonding orbitals, 
$A_{0\sigma}=\frac{1}{\sqrt{2}}(f_{\sigma} - p_{0\sigma})$ and
$B_{0\sigma}=\frac{1}{\sqrt{2}}(f_{\sigma} + p_{0\sigma})$ .The coupling constant is given by
$\lambda \equiv -V_{1}/\hbar\omega$.
The ground states are doubly degenerate 
\begin{align}
|\text{L}\bigl> = &B^{\dagger}_{0\uparrow}B^{\dagger}_{0\downarrow}|\text{vac}\bigl>|\ell\bigl>
\ \ \ \ \text{with}\ |\ell\bigl> = e^{-2\lambda(a-a^{\dagger})}|0\bigl>\label{L}\\
|\text{R}\bigl> = &A^{\dagger}_{0\uparrow}A^{\dagger}_{0\downarrow}|\text{vac}\bigl>|r\bigl>
\ \ \ \ \text{with}\ |r\bigl>    = e^{ 2\lambda(a-a^{\dagger})}|0\bigl>\label{R}
\end{align}
and the binding energy is $E_{\text{PD}} = -4\frac{2}{1+\Lambda^{-1}}\frac{(V_{1})^{2}}{\hbar\omega}$.
In the state $|\text{L}\bigl>$, the bonding state is doubly occupied by 
up and down electrons forming a spin-singlet state and the ionic position is shifted to left, 
while in the state $|\text{R}\bigl>$, the antibonding state is doubly occupied and the ionic position 
is shifted to right.

\begin{figure}
   \begin{center}
      \begin{tabular}{cc}
         \resizebox{40.5mm}{!}{\includegraphics{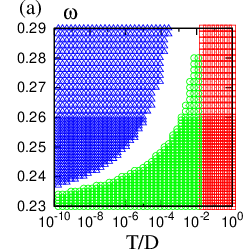}}
         \resizebox{40.5mm}{!}{\includegraphics{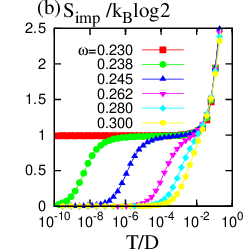}}\\
      \end{tabular}
   \end{center}
   \caption{(Color online)
(a) Entropy of the noninteraction Yu-Anderson model.
(b) Temperature dependence of entropy for various value of phonon frequency $\omega$.
In the panel (a) the red squares are the points where the entropy is larger than $1.1$k$_{\text{B}}\log 2$ 
and the green dots are for $0.9$k$_{\text{B}}\log 2 <$ S$_{\text{imp}} < 1.1$k$_{\text{B}}\log 2$, 
and the blue triangles for S$_{\text{imp}}$ $< 0.1$k$_{\text{B}}\log 2$ 
with the fixed points corresponding to Fig.\ref{site}(c).}
   \label{non_int}
\end{figure}

Kondo effects discussed by YA represent screening process of the $|\text{L}\bigl>$ and $|\text{R}\bigl>$ 
states. In the original model of YA, the localized orbital is not introduced and only scattering processes of 
conduction electrons between the $s$-wave and $p$-wave channels are considered. 
Therefore the original model by YA may be expressed by Fig.\ref{site}(b). 
However in the non-interacting case, $U$=$0$, essential physics of the two models is the same. 
Therefore the model discussed in this Letter spans the Kondo effects of a magnetic impurity to 
the YA type Kondo effects (YAK) due to strong electron-phonon coupling. 
It should be mentioned that the same model as the present one is discussed by Dias da Silva and 
Dagotto in connection with phonon-assisted tunneling in molecular junctions\cite{Dagotto}.

In this Letter, we will show first that YAK are actually realized in the present model by using the NRG and 
then determine the phase diagram in the $U$-$V_{1}$ plane which clarify the essential physics of 
the Kondo effects of a vibrating magnetic ion.

Now we will show results of NRG calculations. Concerning the parameters of the NRG, we keep 
$15,000$ states with $50$ phonon states and use the cutoff $\Lambda$=$3.0$.

\begin{figure}[!ht]
   \begin{center}
      \begin{tabular}{cc}
         \resizebox{40.5mm}{!}{\includegraphics{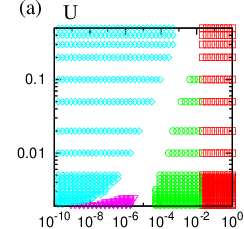}} 
         \resizebox{40.5mm}{!}{\includegraphics{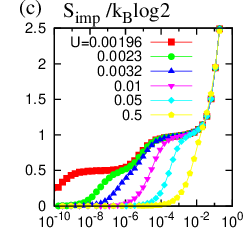}}\\
         \resizebox{40.5mm}{!}{\includegraphics{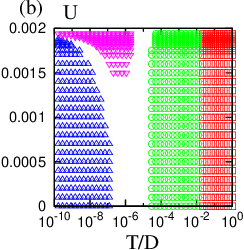}}
         \resizebox{40.5mm}{!}{\includegraphics{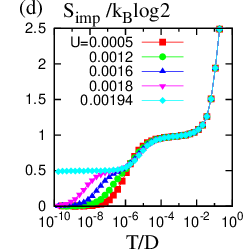}}\\
      \end{tabular}
   \end{center}
   \caption{(Color online)
Effect of Coulomb interaction for the Yu-Anderson model. 
(a) Phase diagram of in $U$-$T$ plane for $U > U_{\text{c}}$, 
where $U_{\text{c}}=0.00194$ is the Coulomb interaction where a 2chK is observed. 
(b) For $U < U_{\text{c}}$.
(c) Temperature dependence of entropy for various $U > U_{\text{c}}$. 
(d) For $U < U_{\text{c}}$.
In the panel (a) and (b) the purple triangles(downward) 
are the points where the entropy is 
$0.4$k$_{\text{B}}\log 2 <$ S$_{\text{imp}} < 0.6$k$_{\text{B}}\log 2$,
and the aqua diamonds for S$_{\text{imp}}$ $< 0.1$k$_{\text{B}}\log 2$ 
with the fixed points corresponding to Fig.\ref{site}(d).}
   \label{int}
\end{figure}

Let us start from the non-interacting case. Fig.\ref{non_int}(b) shows temperature dependence 
of entropy for the impurity part including entropy associated with the phonon degrees of freedom. 
For this example the coupling constants are kept fixed, $V_{0}$=$V_{1}=0.2$ and 
the phonon frequency $\omega$ is varied. When $\omega$ gets smaller more phonons are excited, 
leading to effectively stronger electron-phonon interaction. For $\omega$=$0.300$ the entropy is 
monotonically released as temperature is lowered, which suggests that the system is in the 
RF chain regime. As $\omega$ decreases one can see emergence of 
a plateau in the entropy curve at around k$_{\text{B}} \log 2$.  
This is due to formation of polaron doublet(PD) which correspond to $|\text{L}\bigl>$ 
and $|\text{R}\bigl>$. One can see that entropy release after the plateau for $\omega$=$0.262$, $0.245$, 
and $0.238$ are typical examples of the YAK. To the best of the authors' knowledge 
this is the first NRG calculation for the spinful YAK. When we proceed to the four-site problem 
from the two-site problem
one can see that matrix elements which connect the bonding and antibonding orbitals appear.
Therefore, 
in the present model 
matrix elements between the $|\text{L}\bigl>$ and $|\text{R}\bigl>$ naturally 
appears on this level and YAK are found in a reasonable range of parameters. 
On the contrary, when we use the original YA model, Fig.\ref{site}(b), and assume complete symmetry between 
the $s$- and $p$-wave channels, there is no matrix elements for the mixing. Therefore higher order 
processes like two-phonon processes are required for the YAK to be realized\cite{Matsuura_Miyake}.

When $\omega$ gets smaller, then the binding energy of the PD becomes larger and 
the Kondo temperature of the YAK becomes extremely small. 
Therefore YAK are not seen in the temperature range studied when $\omega$ $<0.235$. 
The panel (a) summarizes the results on the entropy in $\omega$-$T$ plane.  
One can estimate the Kondo temperature of YAK from the middle of the white region.

To see the effects of Coulomb interaction on the YAK, finite $U$ is introduced 
for a fixed $\omega$=$0.245$. The results are shown in Fig.\ref{int}. 
The panels (b) and (d) show that when the Coulomb interaction is weak the behaviors of the YAK do not change. 
However, in a narrow range of $U$ around $U$=$0.00194$ a new plateau appears in the temperature 
dependence of the entropy. The new plateau region is shown by purple triangles(downward) in the panels (a) and (b).
The fact that the entropy at the plateau is (1/2)k$_{\text{B}}\log 2$ suggests that the plateau is due 
to a two-channel Kondo effect(2chK). Actually our analysis of the energy spectrum at this critical $U$ 
confirms that the new fixed point is the 2chK fixed point\cite{cox}.
Through the critical $U_{c}$=$0.00194$ the plateau region of k$_{\text{B}}\log 2$ at higher temperatures remains 
for $U < 0.2$ which is of the order of the binding energy of the PD 
estimated to be $0.24$. We have looked susceptibility of the system and it is confirmed that 
the plateau is due to the PD since there is no enhancement of the susceptibility in this parameter range.

Dias da Silva and Dagotto have found a 2chK behaviors for the set of parameters $V_{0}$=$0.2$, $V_{1}$=$0.1685$, $U$=$0.5$, 
$\omega$=$0.2$\cite{Dagotto}. 
We have confirmed that 
the 2chK behavior is seen for the set of parameters $V_{0}$=$0.2$, $V_{1}$=$0.1646$, $U$=$0.5$, 
and $\omega$=$0.2$. The small difference of $V_{1}$ can be attributed to difference 
in details of numerical calculations.
It is important to figure out the relation between the 2chK fixed point 
in the intermediate coupling region and the one in the close vicinity of the YAK.

\begin{figure}
   \begin{center}
      \resizebox{64mm}{!}{\includegraphics{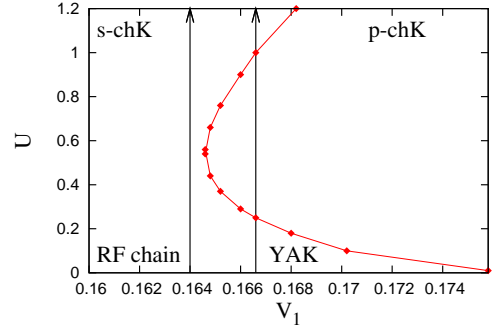}} 
   \end{center}
   \caption{(Color online)The phase boundary between the $s$-chK fixed point and the $p$-chK fixed point.}
   \label{V_1_vs_U}
\end{figure}

Figure.\ref{V_1_vs_U} shows the separatrix on which the 2chK behaviors are observed in the $U$-$V_{1}$ plane 
for $V_{0}$=$0.2$, $\omega$=$0.2$ fixed. When we consider the situation where typical $s$-chK is 
observed the energy spectrum for finite systems at the low energy fixed point shows oscillatory behaviors consistent 
with the schematic Fig.\ref{site}(c). On the other hand when the system shows $p$-chK, 
oscillatory behaviors of the energy spectrum is characterized by Fig.\ref{site}(d). We have found for any set of 
parameters the energy spectrum shows one of these behaviors. The separatrix is the boundary between the two 
behaviors.

The upper left corner of Fig.\ref{V_1_vs_U} corresponds to $s$-chK and the upper right corner to $p$-chK. 
Therefore it is
clear that the boundary in the large $U$ region represents the 2chK whose origin is 
competition between the $s$-wave screening and the phonon assisted $p$-wave screening. The degrees of freedom to be screened 
is the spin of the localized moment on the magnetic impurity.

\begin{figure}
   \begin{center}
      \begin{tabular}{c}
         \resizebox{70mm}{!}{\includegraphics{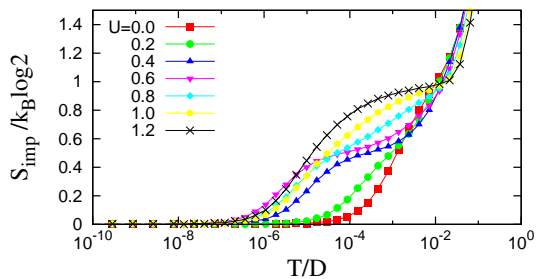}}
      \end{tabular}
   \end{center}
   \caption{(Color online)Temperature dependence of entropy for several parameters along the cut $V_{1}$=$0.164$ in Fig.\ref{V_1_vs_U}.}
   \label{only_s}
\end{figure}

Now we decrease $U$ from the region of the $s$-chK for $V_{1}=0.164$ fixed, Fig.\ref{only_s}.  
For $U=1.2$, we see a typical Kondo behaviors of the entropy. Around $U$=$0.5$, 
in low but finite temperature ranges, 
we see reminiscent of 2chK behaviors which show that the finite temperature properties are influenced by 
the unstable 2chK fixed point.For small $U$ the RF chain behaviors are observed.

\begin{figure}
   \begin{center}
      \begin{tabular}{c}
         \resizebox{70mm}{!}{\includegraphics{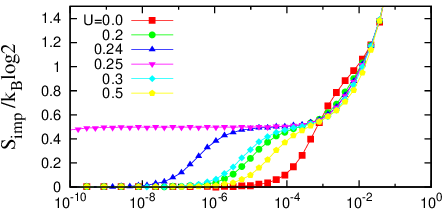}}\\
         \resizebox{70mm}{!}{\includegraphics{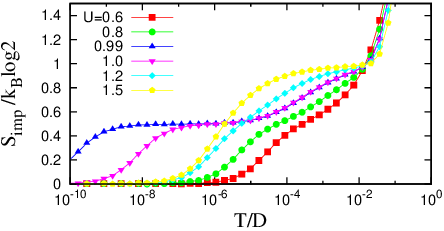}}
      \end{tabular}
   \end{center}
   \caption{(Color online)Temperature dependence of entropy for several $U$ along the cut $V_{1}$=$0.1666$. }
   \label{down_up}
\end{figure}

On the other hand along the cut of $V_{1}$=$0.1666$, Fig.\ref{down_up}, we first observe the $s$-chK 
behaviors for $U$=$1.5$ 
with a shoulder corresponding to k$_{\text{B}}\log 2$ due to the formation of a magnetic moment. 
Then at $U$=$0.99$ the first 2chK behaviors is observed. At $U$=$0.8$ we observe a $p$-chK behavior 
influenced by the 2chK fixed point. As we approach the lower boundary at $U$=$0.25$ we see 
typical behaviors of the 2chK again. For $U$=$0.0$ a shoulder at k$_{\text{B}}\log 2$ is seen which is
due to the PD.
From the phase diagram we have established that the 2chK behaviors 
in the strong correlation regime is continuously connected to the 2chK in the close vicinity of YAK 
in the weak correlation regime.

The latter 2chK is associated with the PD. From the point of view of electron states one component of PD, 
at the two-site level, is the bonding singlet while the other is the antibonding singlet as shown 
in eq.(\ref{L}) and (\ref{R}). These two degrees of freedom may be also described by 
$\psi _{\text{LS}}=\frac{1}{\sqrt{2}}(f^{\dagger}_{\uparrow  }f^{\dagger}_{\downarrow} 
                             + p^{\dagger}_{\uparrow  }p^{\dagger}_{\downarrow})|\text{vac}\bigl>$
and
$\psi _{\text{PS}}=\frac{1}{\sqrt{2}}(f^{\dagger}_{\uparrow  }p^{\dagger}_{\downarrow} 
                             - f^{\dagger}_{\downarrow}p^{\dagger}_{\uparrow  })|\text{vac}\bigl>$
which represent local singlet channel and pair singlet channel. 
For the noninteracting case of the two site problem
they have precisely the same weight resulting in the PD.  
When we attach the $s$-channel and $p$-channel chains, the two local singlets will be extended into chains to 
gain kinetic energy, leading eventually to the same fixed point as the $s$-chK. 
On the other hand when $U$ is finite the pair singlet channel becomes more relevant. 
At the critical point the two channels exactly balance and the 2chK takes place. Important point is that 
concerning the YAK, the degrees of freedom to be screened are the ion displacement described by 
$|\ell\bigl>$ and $|r\bigl>$ and electronic part defines the channels to screen. 
The two different channels are schematically shown by Fig.\ref{site}(c) and (d).  
At this point we would like to comment on the theory by Kusunose and Miyake\cite{Kusunose_Miyake} 
who proposed a possibility of the 2chK for the YAK. They considered that the spin degrees of 
freedom of conduction electrons played the role of screening channels. However, as the wave functions 
$|\text{L}\bigl>$ and $|\text{R}\bigl>$ show, electronic parts of the PD are spin-singlets and their discussion 
is not justified for the spinful YAK.

In conclusion we have derived a generalized Anderson model 
for a vibrating magnetic ion coupled with conduction electrons. 
We have studied the simplest case of $s$-wave localized orbital and 
ion vibrations in one direction. The model shows the $s$-chK and $p$-chK as well as the YAK behaviors. 
The relation among them is clarified by the NRG calculations. 
It is shown that there is a 2chK unstable fixed points in the close vicinity of the YAK in weak correlation regime. 
The boundary is continuously connected with the 2chK behaviors in the strong correlation regime.

The present results suggest that Kondo behaviors, in particular a large mass enhancement, 
may be realized in a wide range of parameters once the vibration frequency of a magnetic ion
is small and the electron-phonon coupling is strong. 
This mechanism of mass enhancement takes place in the weak correlation regime and therefore 
is expected to be robust against magnetic fields. 
In this Letter we have treated only harmonic phonons. However, 
there is a growing consensus that rattling type anharmonic phonons provide 
low energy vibration modes with effectively strong coupling\cite{dalm_ueda,takechi_ueda,hattori_tsunetsugu}. 
It is an interesting future problem to investigate Kondo effect of a vibrating magnetic ion 
in strongly anharmonic potential.

This work is supported by Grant-in-Aid on Innovative Areas "Heavy Electrons"(No. 20100208) 
and also by Scientific Reserch (C) (No.20540347). S.K is supported by JSPS Grant-in-Aid for 
JSPS Fellows 21$\cdot$6752.

\end{document}